\documentclass[preprint]{elsarticle}
\usepackage{amssymb}
\usepackage{amsmath}
\usepackage{amsfonts}
\usepackage{graphicx}
\graphicspath{{figs/}}
\DeclareGraphicsExtensions {.pdf, .jpg, .jpeg, .gif, .png}
\usepackage{color}

\begin{document}

\vspace{5mm}

\title{Variety of scaling laws for DNA thermal denaturation}

\author[1,2,3]{Yulian Honchar}
\ead{julkohon@icmp.lviv.ua}

\author[3,2]{Christian von Ferber}

\author[1,2,3]{Yurij Holovatch}

\affiliation[1]{organization={Institute for Condensed Matter Physics, National Acad. Sci. of Ukraine},
	postcode={79011},
	city={Lviv},
	country={Ukraine}}

\affiliation[2]{organization={$L^4$ Collaboration \& Doctoral College for the Statistical Physics of Complex Systems, Leipzig-Lorraine-Lviv-Coventry},
		country={Europe}}

\affiliation[3]{organization={Centre for Fluid and Complex Systems, Coventry University},
	postcode={CV1 5FB},
	city={Coventry},
	coutry={United Kingdom}}

\date{\today}

\begin{abstract}
 {We discuss possible mechanisms that may impact the
 order of the transition between denaturated and bound DNA states
 and lead to changes in the scaling laws that govern conformational
 properties of DNA strands. To this end, we re-consider the
 Poland-Scheraga model and apply a polymer field theory approach to
 calculate entropic exponents associated with the denaturated loop
 distribution. We discuss in particular variants of this transition
 that may occur due to the properties of the solution and may affect
 the self- and mutual interaction of both single and double strands.
 We find that the effects
 studied significantly influence the strength of the first order
 transition. This is manifest in particular by the changes in the
 scaling laws that govern DNA loop and strand distribution. As a
 quantitative measure of these changes we present the values of
 corresponding scaling exponents. For the $d=4-\varepsilon$ case we
 get corresponding $\varepsilon^4$ expansions and evaluate the
 perturbation theory expansions at space dimension $d=3$
 by means of resummation technique.}
\end{abstract}
\begin{keyword}
DNA denaturation \sep Scaling exponents \sep $\varepsilon$-expansion
\end{keyword}
\maketitle
\medskip

\section{Introduction}\label{I}

In its native state  DNA has a form of a helix that consists of two strands bound together by hydrogen bonds. 
During biological processes involving DNA (such as duplication or transcription) unbinding occurs, phenomenon 
known also as  denaturation, helix-to-coil transition or DNA unzipping \cite{Wartell85,Poland66a,Poland66b}.
An analogue to DNA unwinding in a cell can be observed also {\em in vitro}, in solutions of purified DNA. Already in the middle 50-ies of 
the last century it has been observed that heating of DNA solutions above room temperature results in  cooperative transition 
of the bound helix-structures strands to single strands. Although the mechanism of such unwinding clearly differs from the 
biological protein-mediated process,  ongoing experimental studies of DNA alone are important steps toward understanding much 
more complex phenomenon that occurs {\em in vivo} in a cell \cite{4}. 
The mechanism of such transition 
may be an external pulling force applied to one of DNA strands (mechanical unzipping), changes 
in the pH of the DNA contained solvent (chemical unzipping) or its heating (thermal unzipping) \cite{Wartell85}. The 
scaling laws that govern this last phenomenon are the subject of analysis in our paper. It is our great honour 
to submit this paper to the Physica A special issue in memoriam of Dietrich Stauffer: his contribution to statistical
physics in general and to studies of scaling properties of many-agent interacting systems is hard to be overestimated.

One of important experimental observations of the DNA melting curves, where the fraction of the bound pairs $\theta(T)$ 
is measured as a function of temperature $T$ is their abrupt behaviour \cite{Wartell85,Reiter15}. With an increase of $T$,  $\theta(T)$ manifests a 
jump at certain transition temperature clearly signaling that the DNA thermal denaturation is a first order transition. 
Numerous theoretical approaches to represent the process of DNA thermal denaturation in a two-state Ising-like manner were 
developed. Here, we concentrate on the Poland-Scheraga type description, where the transition is governed by an interplay
of two factors: chain binding energy and configurational entropy \cite{Poland66a,Poland66b,Poland70}. In turn, the entropy 
of the macromolecule in a good solvent
attains a scaling form and this is how the scaling exponents that govern configurational properties of polymer
macromolecules of different topology \cite{Duplantier,Schaefer92} come into play in descriptions of DNA thermal denaturation 
\cite{Kafri00,Kafri02,Baiesi02,Carlon04}. 

Similar as the Ising model suggested to describe ferromagnetism fails if one models a ferromagnet as a 1d 
chain \cite{Ising}, the Poland-Scheraga model suggested to describe the 1st order transition of DNA thermal 
denaturation fails  (predicting the 2nd order scenario) if one models DNA strands as non-interacting random 
walks (RWs) \cite{Poland66a,Poland66b}. With a span of time (see a short overview in the next section) it became clear
that an account of in- and inter-strand interactions plays crucial role and leads to correct picture 
of the transition. In particular the role of interactions between denaturated DNA loop and bound chain
has been analyzed both numerically \cite{Carlon02,Baiesi02} and analytically \cite{Kafri00,Kafri02}. Corresponding analytical calculations
have been performed by field-theoretic approach in $d=4-\varepsilon$ dimensions with 
$\varepsilon^2$ accuracy \cite{Kafri02}. Besides, it was suggested
\cite{Baiesi02} that chain heterogeneity may impact the order of the transition too. 

So far, the origin of such heterogeneity
has been attributed  to effective scaling behaviour differing from that of a self-avoiding walk (SAW). However, it is
well known that depending on temperature, the asymptotic scaling behaviour of a flexible macromolecule belongs 
either to RW ($T=T_\theta$) or to SAW ($T>T_\theta$) universality 
class ($T_\theta$ denoting the $\theta$-point) \cite{ternary}. Therefore, it is tempting to recast  the heterogeneity in scaling
behaviour of a single- and double-stranded chains by studying asymptotic scaling properties of mutually
interacting SAWs and RWs. This is the main goal of our paper. To this end, we 
apply a polymer field theory to calculate entropic exponents associated with the denaturated 
loop distribution. Doing so, besides calculating a set of new scaling exponents that
govern DNA loop and strand distribution for heterogeneous chains we also
get higher precision values of the familiar exponents that have 
been calculated for the homogeneous case \cite{Kafri02}.  The set-up of the paper
is as follows: in the next section \ref{II} we give a short history of
Poland-Scheraga model and discuss the polymer network description for DNA denaturation.
There we obtain scaling relations that enable quantitative analysis of
the order of DNA thermal denaturation transition in terms of co-polymer
star exponents. We obtain perturbation theory expansions for the scaling
exponents and evaluate them in 3d case in section \ref{III}. Conclusions and
outlook are given in section \ref{IV}.

\section{Poland-Scheraga model and polymer network description for DNA denaturation} \label{II}

One of the first models to explain the DNA denaturation as the first order transition was coined in middle-sixties by 
Poland and Scheraga \cite{Poland66a,Poland66b}. Two observations lay at the core of the model:
(i) the breaking of hydrogen bonds between nucleobases is an energy-demanding process, therefore 
the low-$T$ bound state is energetically favoured, and (ii) a high-$T$ unbound state has more 
configurations and hence it is  favoured by entropy. Being simple enough to allow analytic and numerical treatments 
and at the same time capturing main peculiarities of the phenomena involved, the model gave rise to a whole direction 
of studies, see e.g. \cite{Poland70,Reiter15,Kafri00,Kafri02,Baiesi02,Carlon04,Carlon02,Fisher66,Richard04,Berger20,Legrand21}.

\begin{figure}[h!]
	\includegraphics[width=\linewidth]{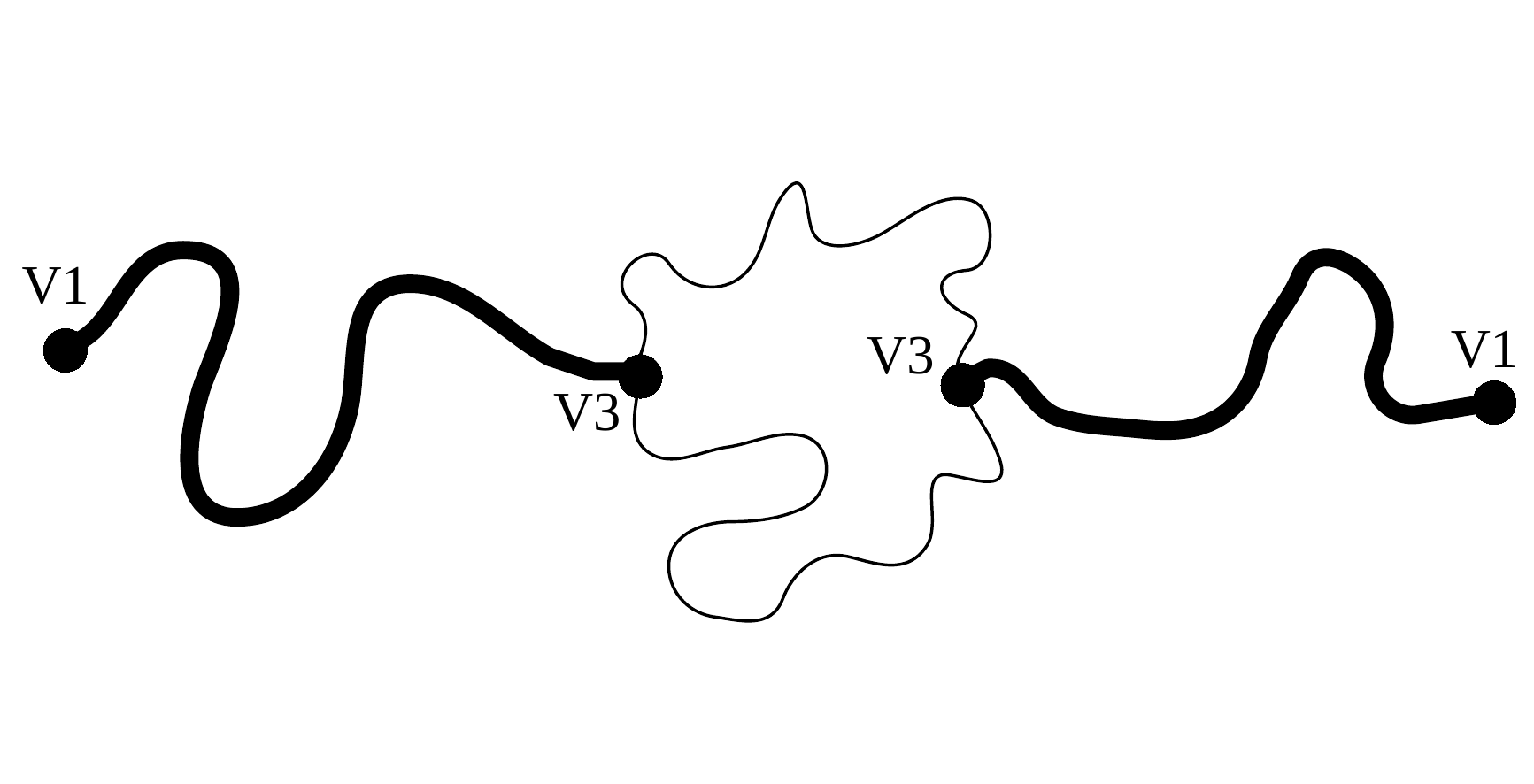}
	\caption{Model of DNA thermal denaturation we consider in this study. A double stranded
	macromolecule with end points $V1$ is disattached (`unzipped') at points $V3$. Resulting
	heterogeneous polymer network consists of two double strands $V1-V3$ (bound nucleobases,  bold lines)
	and a denaturated single-stranded loop  $V3-V3$ (unbound nucleobases, thin lines). Partition function 
	of the loop attains a power-law scaling
	(\ref{00}).} \label{fig1}
\end{figure}

The Poland-Scheraga description, relies on a representation of the partition function ${\cal Z}$ of a 
polymer of $N$ segments,  each segment being in two possible states (bound and unbound monomers, 
see Fig. \ref{fig1}) in a form 
${\cal Z}=x_1^N$, $x_1$ being maximal solution of the equation suggested in~\cite{Lifson64}. In turn, this 
allows to get the order parameter $\theta(T)$ (average
number of ordered, bound pairs in a chain) and to observe different regimes for its temperature $T$ dependence. 
These regimes are triggered by the loop closure exponent $c$ for a single loop, defined as
\begin{equation} \label{00}
{\cal Z}_{\rm loop} \sim \mu^\ell \ell^{-c},
\end{equation}
where $\ell$ is loop length (number of segments) and $\mu$ is a non-universal factor. In particular, for low values of $c$, 
$0 \leq c \leq 1$ the order parameter is continuous function of $T$ smoothly changing between 0 and 1 when 
$T$ decreases from $\infty$ to 0. For larger values of $c$ the order parameter either continuously 
vanishes at $T=T_c$ for $1<c\leq 2$ or disappears abruptly at $T=T_c$ for $c > 2$. 
The last two types of behaviour correspond to the second and first order phase transitions.
 
First attempts to define exponent $c$ analytically let to $1<c<2$ and hence to the second order transition
scenario. In particular, for a simplified model which considers 
a single denaturated loop and does not take into account 
interaction between bound and unbound segments (non-interacting RWs) one may obtain $c$ by 
enumerating walks that return to the origin leading to  $c=1$, $c=3/2$  for $d=2$, $d=3$, correspondingly
\cite{Poland66b}. 
A more general formula $c=d\nu$, with $\nu$ being polymer end-to-end distance scaling exponent was coined 
by Fisher \cite{Fisher66}. In particular, it allows to take into account excluded volume effects for each of 
the segments. Its prediction is $c(d=2)=3/2$,  $c(d=3)\simeq 1.764$ and hence the transition still 
remains the second order. An interaction of a loop with the rest of the chain was taken into account 
in~\cite{Kafri00,Kafri02} by making use of polymer network scaling description 
\cite{Duplantier,Schaefer92}. There, the configurational properties of a homogeneous SAW
polymer  network  with a single denaturated loop were recast in terms of the corresponding 
scaling exponents. The phase transition was found to be of the first order for $d=2$ 
and above, with $c(d=2)=2 + 13/32\simeq 2.41$,  $c(d=3)\simeq 2.115$. Numerical simulations at $d=3$ further 
supported the first order scenario with $c(d=3)= 2.10(4)$ \cite{Carlon02} and  $c(d=3)= 2.18)6)$ \cite{Baiesi02}.  
Effect of possible heterogeneity was partially taken into account 
in~\cite{Baiesi02} by assuming that entropic scaling exponents may differ for different parts
of the network and introducing fit parameters to quantify such difference. 

Let us find the exponent $c$, Eq. (\ref{00}) that governs scaling of a denaturated loop in the simplified
picture of DNA unzipping considering that the macromolecule consists of chains of two different
species, as shown in Fig. \ref{fig1}. The loop is formed by unbound nucleobases, let us take it to be
of `species 1' whereas the two chains $V1-V3$, $V3-V1$ consist of bound nucleobases, `species 2'.
In order to evaluate entropy of a single loop in a network in Fig. \ref{fig1}, we will make use of 
scaling picture for copolymer networks, as suggested in~\cite{Ferber97,Ferber99}. In particular,
the partition function (number of configurations) of a copolymer nework ${\cal G}$ made of $F_1$ chains 
of species 1 and $F_2$ chains of species 2 scales with a mean
size of a single polymer chain $R$ as \cite{Ferber97}:
\begin{equation}\label{1}
{\cal Z_ G}\sim R^{\eta_{\cal G}-F_1\eta_{2,0}-F_2\eta_{0,2}},
\end{equation}
with
\begin{equation}\label{2}
\eta_{\cal G}=-d{\cal L} +\sum_{f_1+f_2\geq 1} n_{f_1,f_2} \eta_{f_1,f_2},
\end{equation}
where $d$ is space dimension, ${\cal L}$ is the number of loops in the network, 
$n_{f_1,f_2}$ is number of vertices where $f_1$ chains of species 1 and $f_2$ 
chains of species 2 meet. Exponents $\eta_{f_1f_2}$ constitute a family of copolymer 
star exponents \cite{Ferber97}. Each of them describes scaling of a copolymer star of corresponding
functionality, made of $f_1$ chains of species 1 and  $f_2$ chains of species 2.
Being universal, they  depend only on space
dimension $d$ and the number of chains $f_1$, $f_2$, as well as three
different types of fixed points (FPs) that govern scaling
behavior \cite{ternary}. These FPs correspond to the cases when
(i) both species 1 and 2 are mutually interacting SAWs, the so-called symmetric
fixed point $S$, (ii) species 1 and 2 are mutually interacting
SAWs and RWs, correspondingly, unsymmetric fixed point $U$ and (iii) both species 1 and 2
are RWs, however there is mutual avoidance interaction between species 1 and 2, 
fixed point $G$. By case
(i) one recovers the homogeneous polymer network, whereas cases (ii) and (iii) present
non-trivial examples of copolymer scaling. Therefore the scaling  properties
of a heterogeneous polymer network made of interacting SAWs and RWs can be reformulated
in terms scaling exponents of co-polymer stars made of two interacting sets of SAWs ($\eta^S_{f_1f_2}$),
of RWs ($\eta^G_{f_1f_2}$) or of a set of SAWs that interacts with RWs ($\eta^U_{f_1f_2}$) \cite{note1}.
Field-theoretical renormalization group
calculations of the above copolymer star exponents resulted in $\varepsilon=4-d$ expansions
which have been obtained successively within $\varepsilon^3$ \cite{Ferber97} 
and $\varepsilon^4$ \cite{Schulte-Frohlinde04} accuracy. 

According to the above, one can distinguish four different cases that
take into account heterogeneity of the network shown in Fig. \ref{fig1} and 
consider mutual avoidance between all SAWs and
RWs. In each of this cases, applying Eq. (\ref{2}) we arrive at the following 
expressions for $\eta_{\cal G}$:
\begin{itemize}
	\item[(i)] {\em{SAW-SAW-SAW:}} both chains $V_1-V_3$ and $V_3-V_3$ are SAWs,
	\begin{equation} \label{2a}
	\eta_{\cal G}= -d + 2 \eta^S_{12}
	\end{equation}	
	\item[(ii)]  {\em{SAW-RW-SAW:}} chains $V_1-V_3$ are SAWs, chains $V_3-V_3$ are RWs;
	\begin{equation} \label{2b}
	\eta_{\cal G}= -d + 2 \eta^U_{12}
	\end{equation}
	\item[(iii)]  {\em{RW-SAW-RW:}} chains $V_1-V_3$ are RWs, chains $V_3-V_3$ are SAWs;
	\begin{equation} \label{2c}
	\eta_{\cal G}= -d + 2 \eta^U_{21}
	\end{equation}
	\item[(iv)] {\em{RW-RW-RW:}} all chains $V_1-V_3$ and $V_3-V_3$ are RWs.
	\begin{equation} \label{2d}
	\eta_{\cal G}= -d + 2 \eta^G_{12}
	\end{equation}
\end{itemize}
In the 'symmetric' case (i) we recover usual homogeneous polymer picture by taking into
account that $\eta^S_{12}=\eta_3$, $\eta_3$ being scaling exponent of the homogeneous
three-leg SAW star \cite{Duplantier,Schaefer92}.

With the above expressions for the heterogeneous co-polymer network exponents $\eta_{\cal G}$ at
hand, it is straightforward to proceed deriving loop closure exponents for each of the cases (i)-(iv).
To this end, following \cite{Kafri00,Carlon04} we generalize expression (\ref{1}) to the case when the network is formed
by chains of different sizes: $R$ for the side chains $V1-V3$ and $r$ for the loop $V3-V3$. Then
the expression for the partition function reads:
\begin{equation}\label{3}
{\cal Z_ G}\sim R^{\eta_{\cal G}-F_1\eta^U_{20}} f(r/R).
\end{equation}
Here $f(x)$ is the scaling function, $F_1$ is the number of SAWs in the network and we have taken into
account that the exponent $\eta^U_{02}=0$ for RWs \cite{note1}. Furthermore, considering the limit $r/R\to 0$
we apply the short-chain expansion \cite{short-chain} and make use of the observation  that for vanishing loop
size the partition function (\ref{3}) should reduce to that of a single (either SAW or RW) chain, 
${\cal Z}_{\rm chain}\sim R^{\eta_{\rm chain}}$ with $\eta_{\rm chain}=\eta^U_{02}=0$ for RW
and $\eta_{\rm chain}=\eta^U_{20}$ for SAW \cite{note1}.
This implies the power-law asymptotics for the scaling function:
\begin{equation}\label{4}
f(x)\sim x^{y}, \hspace{1em}{\rm with}\hspace{1em}
y= \eta_{\cal G}-F_1\eta^U_{20} - \eta_{\rm chain} \, . 
\end{equation}
Indeed, with (\ref{4}) the partition function factorizes as
\begin{equation}\label{5}
{\cal Z_ G}\sim R^{\eta_{\rm chain}}\times r^y \sim {\cal Z}_{\rm chain} {\cal Z}_{\rm loop} \, .
\end{equation}
Comparing Eqs. (\ref{00}) and (\ref{4}) one arrives at the following expression for the
loop closure exponent $c$:
\begin{equation}\label{6}
c= \nu_{\rm loop} [\eta_{\rm chain} -\eta_{\cal G}+F_1\eta^U_{20} ] \, ,
\end{equation}
where $\nu_{\rm loop}$ is the end-to-end distance scaling exponent of the loop
forming chain,  ${\cal Z}_{\rm loop} \sim r^{y} \sim \ell^{\nu_{\rm loop}y}$.

Combining formula (\ref{6}) with the corresponding expressions for $\eta_{\cal G}$ of
four different heterogeneous networks formed by mutually interacting SAWs and RWs,
Eqs. (\ref{2a})--(\ref{2d}) we arrive at the following loop closure exponents in each
of these networks, in the following denoted as $c_1$--$c_4$:
\begin{eqnarray} \label{7a}
&\text{\it SAW-SAW-SAW:}& c_1 = \nu_{\rm SAW} (3\eta^S_{20} + d - 2\eta^S_{12})\, , \\ \label{7b}
&\text{\it SAW-RW-SAW:}& c_2 =  \nu_{\rm RW} (\eta^S_{20} + d - 2\eta^U_{12})\, , \\ \label{7c}
&\text{\it  RW-SAW-RW:}& c_3 =  \nu_{\rm SAW} (2\eta^S_{20} + d - 2\eta^U_{21})\, . \\ \label{7d}
&\text{\it  RW-RW-RW:}& c_4 = \nu_{\rm RW} (d - 2\eta^G_{21})\, ,
 \end{eqnarray}
where $\nu_{\rm RW}=1/2$ and $\nu_{\rm SAW}$ are the end-to-end mean distance exponents
for RW and SAW, correspondingly. By (\ref{00}), each of the above expressions govern 
the loop closure in the heterogeneous network and therefore defines the order of 
the phase transition of DNA thermal denaturation in the frames of the Poland-Scheraga model.
In the following section we will evaluate these expressions for the case $d=3$.

\section{$\varepsilon$-expansion and its resummation}\label{III}

Scaling relations (\ref{7a})--(\ref{7d}) express exponents
$c_i$ in terms of the familiar co-polymer star exponents $\eta_{f_1f_2}$ \cite{Ferber97}. 
The latter have been calculated by means of field-theoretic renormalization
group approach and are currently available in $\varepsilon=4-d$-expansion up to order
$\varepsilon^4$ \cite{Schulte-Frohlinde04}. Completing these expansions by
familiar $\varepsilon$-expansion for the exponent  $\nu_{\rm SAW}$, see e.g.
\cite{Kleinert01}, one readily gets the corresponding expansions for exponents
$c_i$:
\begin{eqnarray} \nonumber
c_1=2+1/8\,\varepsilon+{\frac {5}{256}}\,{\varepsilon}^{2}+ \left(
-{\frac {87}{ 4096}}+{\frac {3}{512}}\,\zeta  \left( 3 \right)
\right) {\varepsilon}^{ 3}+
\\ \label{8a}
\left( -{\frac
	{3547}{262144}}+{\frac {903}{16384}}\,\zeta
\left( 3 \right) +{\frac {1}{20480}}\,{\pi }^{4}-{\frac {1815}{2048}}
\,\zeta  \left( 5 \right)  \right) {\varepsilon}^{4} ,
\end{eqnarray}
\begin{eqnarray} \nonumber
c_2 &=& 2+1/8\,\varepsilon+ \left( {\frac {17}{256}}-{\frac
	{21}{64}}\,\zeta
\left( 3 \right)  \right) {\varepsilon}^{2}+ \\  && \nonumber
\left( -{\frac {63}{512}}\,
\zeta  \left( 3 \right) -{\frac {39}{4096}} \right)
{\varepsilon}^{3}+
\Big( -{\frac {3015}{262144}}+  \\ \label{8b}
 && {\frac {105}{16384}}\,\zeta  \left( 3
\right) +
{\frac {1185}{2048}}\,\zeta  \left( 5 \right) -{\frac {21}{
		20480}}\,{\pi }^{4} \Big) {\varepsilon}^{4} ,
\end{eqnarray}
\begin{eqnarray} \nonumber
c_3 &=& 2+3/4\,\varepsilon+ \left( {\frac {21}{128}}-{\frac
	{21}{32}}\,\zeta
\left( 3 \right)  \right) {\varepsilon}^{2} + \\ \nonumber && \left( {\frac {157}{2048}}-
{\frac {27}{128}}\,\zeta  \left( 3 \right)  \right)
{\varepsilon}^{3}+
\Big( {\frac {4125}{131072}}-{\frac {165}{1024}}\,\zeta  \left( 3
\right) - \\ \label{8c} && {\frac {11}{10240}}\,{\pi }^{4}-{\frac {465}{1024}}\,\zeta
\left( 5 \right)  \Big) {\varepsilon}^{4} ,
\end{eqnarray}
\begin{equation} \label{8d}
c_4 = 2+1/2\,\varepsilon \, ,
\end{equation}
where $\zeta(x)$ is Riemann $\zeta$-function. 
Note that because of $\eta^G_{21}=\eta^G_{12}=-1$ \cite{Ferber97}
the corresponding analytic expression for $c_4$ is exact and contains
only linear term.

Perturbative renormalization group expansions have zero radius of convergence and are 
asymptotic at best \cite{Kleinert01,Zinn-Justin89}. Special resummation procedures are used to restore their convergence
and to get reliable numerical estimates on their basis. Below we will make use of the 
Borel resummation refined by conformal mapping \cite{LeGuillou80} which is known to be a powerful
tool in analysis of $\varepsilon$-expansions. In general, the method is applied 
to the function in form of a series expansion:
 \begin{equation} \label{8}
 \upsilon(x)=\sum\limits_{n=0} c_n x^n ,
 \end{equation}
with known asymptotics of the coefficients $c_n$:
 \begin{equation}\label{ass}
\lim_{n\to \infty} c_n \sim n!(-a)^n \, . 
 \end{equation}
The Borel sum associated with (\ref{8}) is used to mitigate the factorial growth of the expansion coefficients
 \begin{equation} \label{Borel1}
  B(x)=\sum\limits_{n=0} \frac{c_n x^n}{n!}\, .
 \end{equation}
 The resulting series is supposed to converge, in the complex $x$-plane, inside a circle of radius $1/a$
 (cf. Eq. (\ref{ass}), where $x=-1/a$ is the singularity of $B(x)$ closest to the origin.
 Then using the definition of the $\Gamma$-function one can rewrite Eq. (\ref{8})
 as 
 \begin{equation} \label{Borel2}
\upsilon(x)  =\sum\limits_{n=0} \frac{c_n x^n}{n!} \int_{0}^{\infty} dt \, e^{-t} t^n \, .
 \end{equation}
 Interchanging summation and integration in Eq. (\ref{Borel2}) leads to the definition of the Borel 
transform of $\upsilon(x)$ as
\begin{equation} \label{Borel3}
 \upsilon_B(x)= \int_{0}^{\infty} dt \, e^{-t} \upsilon(xt)\, .
 \end{equation}
 In order to perform the integral in (\ref{Borel3}) on the whole real
positive semiaxis, one has to find an analytic continuation of $B(x)$.
To this end, assuming that singularities in $B(x)$ lie on the negative 
semiaxis with the singularity closest to the origin located at $1/a$, 
and that the function $B(x)$ is analytical in the complex plane 
excluding the part of real axis $(-\infty;-\frac{1}{a})$, one passes to new 
variables $\omega$ conformally mapping the cut plane onto a disk of radius 1:
\begin{equation} \label{conf}
x=\frac{4}{a}\frac{\omega}{(1-\omega)^2}
\end{equation}
The procedure is further refined by introducing two additional fit parameters $b$ and $\alpha$.
The first one is introduced substituting the factorial $n!$ in Eq. (\ref{ass})  by the Euler gamma-function $\Gamma(n+b+1)$ and inserting 
an additional factor $t^b$ into the integral (\ref{Borel2}). The fit parameter $\alpha$ is introduced to weaken the singularity
by multiplying the expression under the integral by $(1-\omega)^\alpha$. The expression for the resummed function reads:
\begin{equation}\label{res}
\upsilon_R(x)=\sum_n d_n(\alpha,a,b) \int_0^\infty dt \, t^b e^{-t}\frac{\omega^n (xt)}{(1-\omega (t)^{\alpha}} \ .
\end{equation}
Explicit form of the coefficients $d_n(\alpha,a,b)$ is found on the base of known expansion coefficients
$c_n$ in Eq. (\ref{8}). In practice, procedure  (\ref{res})  is applied to the truncated series (\ref{8}), which
is known up to order $L$. Let us denote the value of the resummed truncated function at given fixed $x$ by $\upsilon_R^{(L)}$.
Ideally, such value (that usually corresponds to certain physical observable) should not depend on resummation parameters $\alpha$ and $b$. 
To eliminate such dependence, for each perturbation theory order $L$ one choses optimal values of $\alpha_{opt}^{(L)}, b_{opt}^{(L)}$ which satisfy
condition of minimal sensitivity \cite{Delamotte}:
\begin{equation}
\frac{\partial \upsilon_R^{(L)}(b,\alpha)}{\partial b}\arrowvert_{b_{opt}^{(L)},\alpha_{opt}^{(L)}}=
\frac{\partial  \upsilon_R^{(L)}(b,\alpha)}{\partial \alpha}\arrowvert_{b_{opt}^{(L)},\alpha_{opt}^{(L)}}=0 \, .
\end{equation}
In this way, a set of optimal values $(b,\alpha)^{L}$ is obtained for every perturbation theory order $L$. Out of these points one has 
to choose those that ensure the fastest converge by minimizing values:
\begin{equation}
\upsilon_R^{(L+1)}(b^{(L+1)},\alpha^{(L+1)})-\upsilon_R^{(L)}(b^{(L)},\alpha^{(L)})\, .
\end{equation}

The above described procedure has been applied to obtain the results discussed below.
As we have checked by explicit calculations, the resummation method described above does not lead to consistent results 
 when directly applied
to series (\ref{8a})--(\ref{8c}) at $\varepsilon=1$. This may serve as an evidence of their Borel-nonsummability. Another obvious way to 
get numerical estimates of the exponents $c_i$ is to resum series for the exponents that enter right-hand 
sides of scaling relations (\ref{7a})--(\ref{7c}) and then use these relations to evaluate $c_i$. The resummed values
of exponents $\nu$, $\eta_{f_1f_2}$ are given in Table \ref{tab1} at $d=3$ in different orders of perturbation theory. As one 
can see from the Table, the resummation restores convergence of the $\varepsilon$-expansion for the exponents
and leads to reliable numerical estimates. As a  benchmark one can use the state-of-the art estimate for the 
$\nu_{\rm SAW}$ exponent obtained via conformal bootstrap $\nu_{\rm SAW}(d=3)=0.5877(12)$ \cite{Shimada16} 
and MC simulations $\nu_{\rm SAW}(d=3)=0.5875970(4)$ \cite{Clisby16}
which reasonably well 
compare with our result $0.585(3)$ as given in the Table. Note that $\varepsilon$-expansion for the $\nu_{\rm SAW}$
exponent is currently available with a record $\varepsilon^7$ accuracy and the most recent hypergeometric-Meijer
resummation of this series lead to an estimate $\nu_{\rm SAW}(d=3)=0.58770(17)$ \cite{Shalaby20}.

\begin{table}
\begin{center}
	\begin{tabular}{|l|cccc|}
	\hline
		&{$I$}&{$II$}&{$III$}&{$IV$}\\
		\hline
		$\nu_{SAW}$&0.54(3)&0.56(2)&0.582(8)&0.585(3)\\
		$\eta^S_{20}$&-0.25(6) &-0.292(4) &-0.289 (5)&-0.276(3)\\
		$\eta^S_{12}$ &-0.75(4)&-0.77(2)&-0.75(1)&-0.743(5)\\
		$\eta^U_{12}$&-0.75(8)&-0.82(5)&-0.77(3)&-0.795(5)\\
		$\eta^U_{21}$&-1.(2)&-0.9(8)&-0.95(8)&-0.98(3)\\
		\hline
	\end{tabular}
		\caption{Scaling exponents  $\nu$, $\eta_{f_1f_2}$ obtained at $d=3$ by resummation of 
		$\varepsilon$-expansion in different orders of perturbation theory. \label{tab1}}
\end{center}
\end{table}

\begin{figure}
	\includegraphics[width=0.9\linewidth]{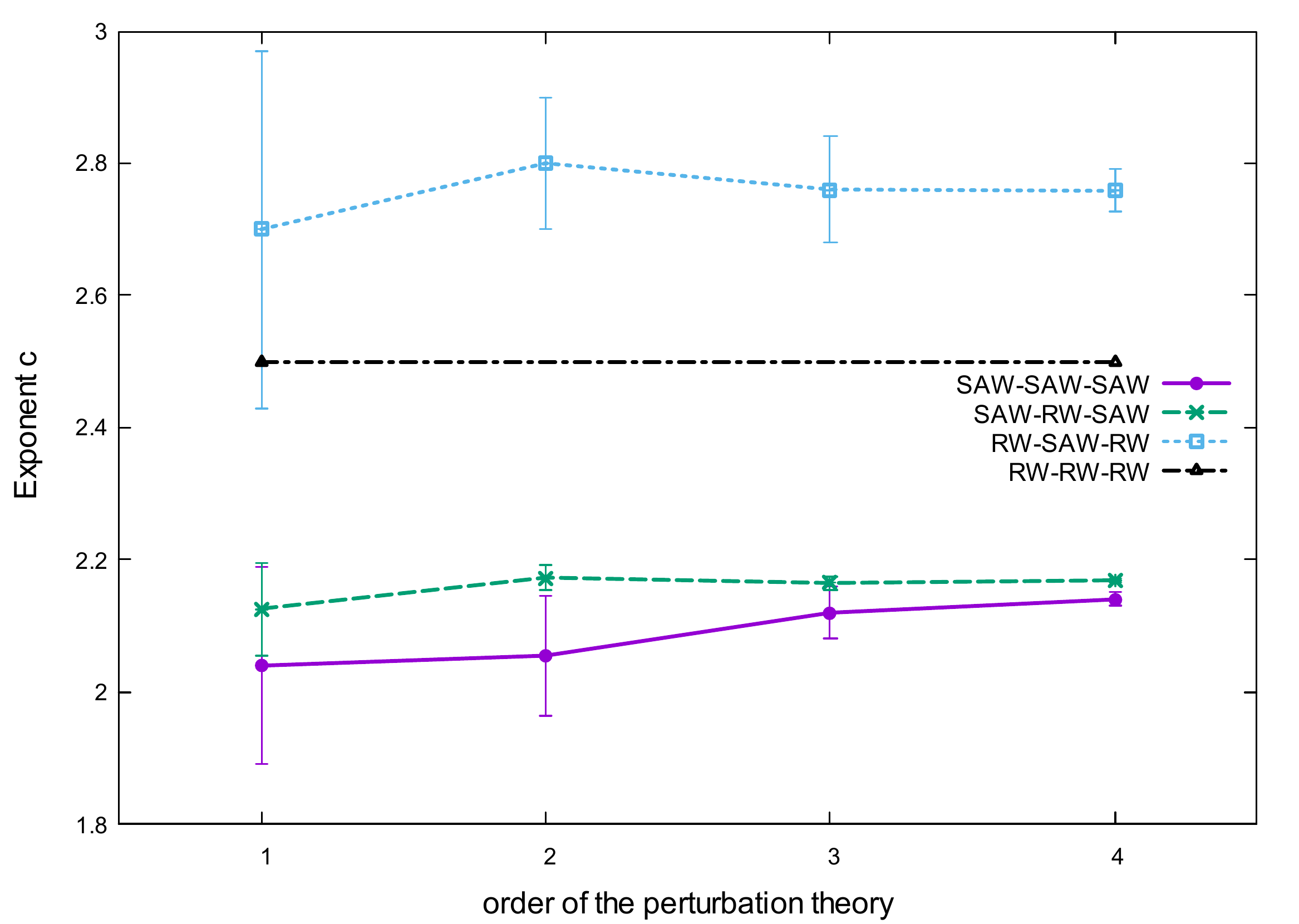}
	\caption{Loop closure exponents $c_i$ of the heterogeneous co-polymer network of interacting SAWs and RWs
	(as shown in Fig. \ref{fig1}) in different orders of the perturbation theory.} \label{fig2}
\end{figure}

\begin{table}
\begin{center}
	\begin{tabular}{|l|cccc|}
		\hline
		
		&{$I$}&{$II$}&{$III$}&{$IV$}\\
		\hline
		$c_1$ (SAW-SAW-SAW) &2.04 (15)&2.05 (9)&2.12 (4)&2.147 (9)\\
		$c_2$ (SAW-RW-SAW)&2.12 (7)&2.17 (2)&2.16 (1)&2.169 (4)\\
		$c_3$ (RW-SAW-RW)&2.7 (3)&2.8 (1)&2.76 (8)&2.76 (3)\\
		$c_4$ (RW-RW-RW)&2.5&2.5&2.5&2.5\\
		\hline
	\end{tabular}
\end{center}
\caption{Loop closure exponents $c_i$ (\ref{7a})--(\ref{7d}) in different orders of perturbation theory
for different combinations of interacting SAWs and RWs. \label{tab2} }
\end{table}

Using numerical estimates for the exponents $\nu$, $\eta_{f_1f_2}$ one can evaluate loop closure exponents $c_i$ 
in successive orders of the $\varepsilon$-expansion. The results are given in Table \ref{tab2} and further displayed 
in Fig. \ref{fig2}. One observes that the numbers nicely converge with an increase of the perturbation theory order
and thus provide reliable numerical estimates for the loop closure exponent in the case of heterogeneus co-polymer
network. These results will be further discussed in the next section.

\section{Conclusions}\label{IV}

Influence of possible heterogeneity in entropic scaling exponents of bound and denaturated DNA strands
on the loop closure exponent $c$ is manifest by an interplay of two factors. On the one hand, the number of configurations
of a denaturated loop, cf. Fig. \ref{fig1}, is influenced by the loop self-avoidance interactions
(the number is larger for the RW loop and smaller of the SAW one). On the other hand, the number of loop configurations is
restricted by the side chains. Calculations presented here give a reliable way to judge about the values of
exponents $c_i$ for different heterogeneity conditions and hence to judge about the order of
DNA thermal denaturation transition. Our analysis is grounded on the field theory of co-polymer
networks \cite{Ferber97,Ferber99}. By scaling relations (\ref{7a})--(\ref{7d}) we connect loop closure
exponents $c_i$
to scaling exponents $\eta_{f1f2}$ that govern entropic properties of co-polymer stars made by mutually interacting 
sets of SAWs and RWs. Using powerful resummation technique, we resum $\varepsilon^4$ expansions 
for these exponents and evaluate them at space dimension $d=3$. Numerical results for the
exponents $c_i$ are listed in Table \ref{tab2} and shown in Fig. \ref{fig2}. As one can see,
the effects of heterogeneity significantly influence the strength of the first order
 transition (the exponent $c$ increases in comparison to the usual homogeneous SAW case).
 
Details of our calculations together with analysis of the 2d case will be a subject of a separate publication \cite{Honchar21}.
We acknowledge useful discussions with Maxym Dudka and Ralph Kenna. 
This work was supported in part by the National Academy of Sciences of Ukraine, project KPKBK 6541230.

\end{document}